# Instability and dynamics of volatile thin films

Hangjie Ji*
*Department of Mathematics, University of California, Los Angeles*
*Los Angeles, CA 90095, USA.*

Thomas P. Witelski†
*Department of Mathematics, Duke University*
*Durham, NC 27708-0320, USA.*
(Dated: August 31, 2017)

Volatile viscous fluids on partially-wetting solid substrates can exhibit interesting interfacial instabilities and pattern formation. We study the dynamics of vapor condensation and fluid evaporation governed by a one-sided model in a low Reynolds number lubrication approximation incorporating surface tension, intermolecular effects and evaporative fluxes. Parameter ranges for evaporation-dominated and condensation-dominated regimes and a critical case are identified. Interfacial instabilities driven by the competition between the disjoining pressure and evaporative effects are studied via linear stability analysis. Transient pattern formation in nearly-flat evolving films in the critical case is investigated. In the weak evaporation limit unstable modes of finite amplitude non-uniform steady states lead to rich droplet dynamics, including flattening, symmetry breaking, and droplet merging. Numerical simulations show long time behaviors leading to evaporation or condensation are sensitive to transitions between film-wise and drop-wise dynamics.

## I. INTRODUCTION

Pattern formation of thin liquid films has been studied in many experimental and theoretical works [44, 49] and has important connections to many engineering applications of coating flows. Morphological changes like localized thinning, rupture, coarsening and droplet dynamics can arise from instabilities caused by the interaction of various physical influences. In the limit of low Reynolds number, the governing Navier Stokes equations can be reduced to a thin film equation for the evolution of the thickness of the fluid layer on a solid substrate. Lubrication models for free surface flows of these thin layers of viscous fluids have been studied extensively in many contexts [15, 31, 32, 35].

In this work, we will take a closer look at a thin film model for a volatile viscous fluid on a solid at a fixed temperature, where the temperature difference with the surrounding vapor can drive evaporation or condensation. Specifically, the aim of this paper is to understand the forms of dynamics that can result from the interaction of substrate wettability and evaporation/condensation of the fluid. Stability analysis will show both significant transient dynamics and strong dependence on initial conditions can occur as the fluid layer evolves between drop-like and film-like states.

Wetting properties of solid substrates can qualitatively affect the behavior of ultrathin layers (< 100 nm) of fluids. While hydrophilic or wetting materials encourage the spreading of fluids to form more uniform films, hydrophobic or non-wetting solids repel the fluid and favor break-up into structures that ultimately leads to arrays of droplets [9, 38, 45]. Generally called "dewetting" phenomenon, these interactions are driven by van der Waals forces between the solid and fluid and are modeled by a disjoining pressure as a contribution to the dynamic pressure in the thin film equation. Early stages of the dynamics involve instabilities of flat films [54] leading to finite-time rupture [55, 56] and later growth of dry spots [20, 25, 30, 38, 45] and coarsening dynamics of interacting meta-stable droplets [22]. These studies have addressed how spatial structures in the films develop subject to the overall

---

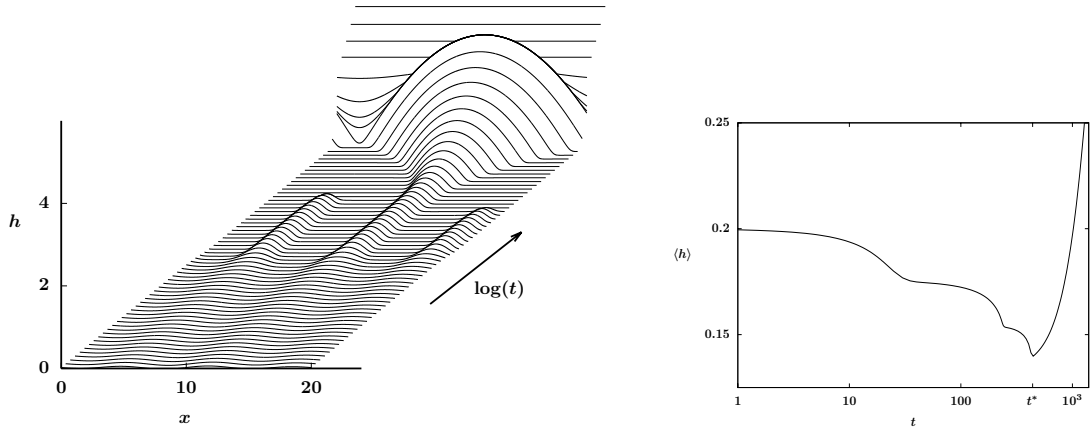

FIG. 1. (Left) A numerical simulation of dewetting of a volatile thin film starting from typical initial conditions: early-stage dewetting yielding three meta-stable droplets, subsequent collapse of the left and right droplets occurs while the center droplet gradually grows, followed by filmwise condensation after the substrate is flooded. (Right) A plot showing the average thickness $\langle h \rangle$ decreasing in the early regime for $0 \leq t < t^* \approx 440$ corresponding to dewetting and evaporation, with later condensation yielding increasing average thickness for $t > t^*$.

conservation of mass for non-volatile fluids. There has been less work on how change of mass from slow evaporation or condensation interacts with dewetting behavior [5, 20, 48].

The study of thin films subject to fluid evaporation and vapor condensation is especially important for systems like precorneal tear films [13] thermal management [29] and drying paint layers [19, 42]. In particular, since dropwise condensation on hydrophobic surfaces is more energy efficient compared to the filmwise condensation mode [39], many studies have considered the influence of the surface wettability on the dynamics [18, 40, 52, 53].

The basic physics of thin films with phase changes has many applications in many systems in engineering and other physical systems [51]. Describing the evaporative mass flux across the fluid-vapor interface generally involves the influence of both liquid phase and the diffusion of the vapor, as in the two-sided model in [47]. Simplified one-sided models are obtained by decoupling the dynamics of the fluid from that of the vapor concentration field. In 1988 Burelbach et al. [14] proposed a one-sided model by assuming that the density, viscosity and thermal conductivity in the vapor are negligible compared to those in the fluid. Following their work, Oron and Bankoff [33, 34] investigated the dynamics of a condensing thin liquid film using an evaporative mass flux only dependent on film thickness and neglected effects like thermocapillarity and vapor thrust. A model was derived by Ajaev and Homsy [1–3] that incorporates thermal effects, surface tension and disjoining pressure in the evaporative flux. In all of these models the dynamics of the vapor phase are not included except through a boundary condition at the vapor-fluid interface. Related problems on evaporating liquid droplets associated with moving contact lines were also studied [1, 6, 7]. For a thorough review on the modeling and numerical studies of volatile thin films, see [4, 15].

While some results are available for the stability and dynamics of volatile polar thin films on partially wettable solid substrates with various forms of evaporative fluxes [1, 2, 36, 43], the nonlinear stability and pattern formation of thin films undergoing phase change on a hydrophobic substrate still needs further investigation. The competition between the evaporation/condensation effects and the disjoining pressure is the focus of our paper. We will show that interfacial instabilities driven by the disjoining pressure can yield transient pattern formation under weak evaporation. Balances between the evaporative effects, intermolecular forces and the surface tension produce a family of spatially uniform or periodic steady states whose instabilities lead to interesting transitions between drop-wise and film-wise dynamics. An example of the complicated dynamics involving film dewetting,

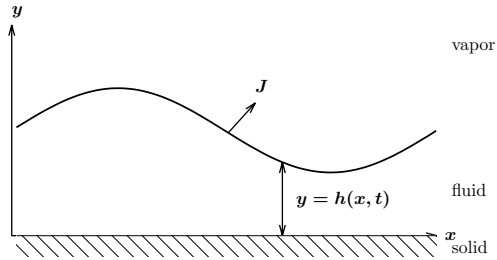

FIG. 2. Schematic representation of a two-dimensional thin fluid film on a uniformaly heated/cooled solid substrate where $J$ represents the evaporative/condensing mass flux.

evaporation, and condensation are shown in Fig. 1.

The structure of this paper is as follows. In section II the model for volatile thin films on a partially-wetting substrate is formulated. In section III spatially uniform steady states will be discussed and the related transient instabilities with respect to spatial perturbations will be investigated via linear stability analysis. Spatially periodic steady states representing droplets will be presented in section IV, and their stability will be analyzed in section V. Numerical simulations of the model and comparison with the analytical predictions for systems with small numbers of droplets are given in section VI and a discussion of remaining open questions is given in section VII.

## II. MODEL FORMULATION

We consider a two-dimensional thin fluid film spreading over a uniformly heated or cooled solid substrate (shown in Fig. 2). Under the long wavelength approximation and in the limit of the low Reynolds number, the classic thin film equation can be derived from the full Navier-Stokes equations [35]. Following [3] we consider an evaporating/condensing thin film equation by including an additional mass flux term to a non-evaporative thin film model. The resulting governing equation in nondimensional form for viscous coating flows is given by a one-dimensional fourth-order nonlinear partial differential equation for the thickness $h$ of the fluid layer over a periodic domain $0 \leq x \leq L$,

$$\frac{\partial h}{\partial t} = \frac{\partial}{\partial x}\left(h^3 \frac{\partial p}{\partial x}\right) - J, \tag{1a}$$

where the evaporative mass flux is given by

$$J(h) = \frac{\beta p(h)}{h + K_0}, \tag{1b}$$

where the two dimensionless parameters $K_0 > 0$ and $\beta > 0$ characterize the effects of phase change kinetics at the interface [5]. Here $K_0$ relates to the thermal resistance of the solid substrate, and $\beta$ scales the dynamic pressure in the evaporation/condensation effects. The influences of surface tension [31] and wetting properties of the substrate are both incorporated into the hydrodynamic pressure $p(h)$ by the linearized curvature $h_{xx}$ and disjoining pressure $\Pi(h)$, respectively,

$$p(h) = \Pi(h) - \frac{\partial^2 h}{\partial x^2}. \tag{1c}$$

Starting from positive and finite-mass initial data $h_0(x) > 0$ at time $t = 0$, the dynamics of the model (1) is governed by the interactions between the mass-conserving and evaporative fluxes [48].

For $\beta = 0$, the governing equation (1) reduces to the non-evaporative equation for the thickness $h$ of the fluid layer

$$\frac{\partial h}{\partial t} = \frac{\partial}{\partial x}\left(h^3 \frac{\partial}{\partial x}\left[\Pi(h) - \frac{\partial^2 h}{\partial x^2}\right]\right). \qquad (2)$$

Different forms of $\Pi(h)$ have been used in the literature to characterize the hydrophobic or hydrophilic properties of the substrate. For instance, the form $\Pi(h) = A/h^3$ (where $A$ is called a Hamaker constant) represents [23] the disjoining intermolecular forces. For the case $A < 0$ this $\Pi(h)$ corresponds to the wetting property of the substrate and acts as a stabilizing conjoining force in the dynamics. It has been shown [54] that instabilities caused by this form of $\Pi(h)$ with $A > 0$ can lead to finite-time rupture, that is, the film thickness $h$ approaches zero at an isolated point as a critical time is approached. The dynamics of these thin film singularities were then studied by [55, 56]. Later a physically-motivated regularized form of the pressure

$$\tilde{\Pi}(h) = \frac{A}{h^3}\left(1 - \frac{\epsilon}{h}\right) \qquad (3)$$

with $A > 0$ was introduced where the combined influences from both the attractive van der Waals forces and short range Born repulsion provide a positive $O(\epsilon)$ lower bound for the film thickness [11, 33, 34, 42].

Various forms of evaporation loss or condensation source terms have been used in the literature. In [8, 14] a simple form for the evaporative mass flux term is taken

$$J(h) = \frac{E_0}{h + K_0},$$

where the dimensionless evaporation number $E_0$ distinguishes the evaporation case with $E_0 > 0$ from the condensation case with $E_0 < 0$. Here $E_0$ represents the scaled difference of the temperature of the heated or cooled solid substrate and the liquid saturation temperature. In the evaporation model for a thin film on a wetting substrate derived by Ajaev and Homsy [1–3] the evaporative flux takes the form of

$$J(h) = \frac{\beta(Ah^{-3} - h_{xx}) + E_0}{h + K_0}, \qquad (4)$$

where both the Kelvin and Clapeyron effects [51] on the vapor pressure are included, and jump conditions in temperature and pressure are imposed at the fluid-vapor interface. The nondimensional parameters $K_0 > 0$ and $\beta > 0$ characterize the relative importance of evaporation kinetics and the change in liquid pressure. In [43, 50] a different form of the evaporative flux was derived by assuming that the flux is proportional to the difference between the chemical potential of the vapor and the liquid.

In order to capture the dynamics of the dewetting films, we adopt the regularized disjoining pressure with combined forces of attractive and repulsive intermolecular forces (3). Moreover, it is observed that adding a constant to the $\Pi(h)$ term will not change the form of the conservative term due to the gradient of the pressure in the PDE (1a). Therefore we define a generalized disjoining pressure $\Pi(h)$,

$$\Pi(h) = \tilde{\Pi}(h) - P_0 \quad \text{with} \quad \tilde{\Pi}(h) = \frac{A}{h^3}\left(1 - \frac{\epsilon}{h}\right), \qquad (5)$$

including a constant $P_0$ pressure offset. This pressure can be expressed as

$$P_0 = (T^* - T_0)/\beta + P_v, \qquad (6)$$

where $P_v$ is the non-dimensional vapor pressure, $T_0$ is the temperature of the substrate and $T^*$ is the liquid saturation temperature. Specifically, $P_0 < 0$ indicates that the vapor phase is undersaturated and $P_0 > 0$ is for supersaturated vapor. With this form of $\Pi(h)$ the pressure term in the evaporative flux (1b) is consistent with that in the conservative term. We will use $A > 0$ in $\tilde{\Pi}(h)$ to identify the hydrophobic property of the substrate, and $\epsilon > 0$ describes the scale of the film where the attractive and repulsive intermolecular forces are dominant. For the rest of the paper, we will use $A = \epsilon^2$ following other studies of dewetting films [11, 33, 35].

An important implication of the form of the evaporative flux (1b) is that the local condensation or evaporation depends on three factors, the film thickness, surface tension, and the pressure $P_0$. For example, evaporation may occur from negatively curved portions of the film ($h_{xx} < 0$) even when the vapor is oversaturated with $P_0 > 0$; and a positively curved portion of the film ($h_{xx} > 0$) may condense when a negative $P_0$ is present if the curvature is large enough. This observation was also commented on in [43].

It is convenient to write the generalized potential as the integral of $\Pi(h)$,

$$U(h) = \int \Pi(h) \ dh = -\frac{\epsilon^2}{2h^2} + \frac{\epsilon^3}{3h^3} - P_0 h. \qquad (7)$$

This form combines the intermolecular forces between the fluid and the solid substrate, and the evaporation/condensation effects through the last term. Following results from [48], we define the energy functional that governs the dynamics of (1) as

$$\mathcal{E}[h] = \int_0^L \frac{1}{2} \left( \frac{\partial h}{\partial x} \right)^2 + U(h) \ dx. \qquad (8)$$

A direct calculation of the time derivative of the functional (8) leads to the dissipation of energy

$$\frac{d\mathcal{E}}{dt} = -\left( \int_0^L h^3 \left( \frac{\partial p}{\partial x} \right)^2 \ dx + \beta \int_0^L \frac{p^2}{h + K_0} \ dx \right) \leq 0. \qquad (9)$$

This observation will be used for a short proof that evaporation can never overcome the disjoining pressure in Appendix A; this behavior is physically expected for adsorbed films, but is not guaranteed for all mathematical forms of evaporative fluxes [24].

Since the contribution from each integral in (9) is nonnegative, at an equilibrium state each integral must equal to zero independently. From the first integral one obtains $\partial p / \partial x = 0$, which indicates that $p$ is a constant over the domain. For $\beta > 0$ the second integral in (9) being zero then leads to $p \equiv 0$. Whether equilibria exist as drops or uniform films depends on the pressure offset $P_0$; our analysis of the dynamics will be partitioned into regimes determined by the sets of equilibrium solutions.

The total mass of the film is of key interest and its rate of change due to the evaporative term can be obtained by integrating equation (1a) and incorporating the periodic boundary conditions,

$$\mathcal{M}(t) = \int_0^L h \ dx, \qquad \frac{d\mathcal{M}}{dt} = -\beta \int_0^L \frac{p}{h + K_0} \ dx. \qquad (10)$$

Using (1c) and performing an integration by parts, we get

$$\frac{d\mathcal{M}}{dt} = \beta \left( \int_0^L \frac{h_x^2}{(h + K_0)^2} \ dx - \int_0^L \frac{\Pi(h)}{h + K_0} \ dx \right). \qquad (11)$$

For the special case with $\beta = 0$ the mass is conserved and no evaporation or condensation occurs (2). In other cases, the increase or decrease of the mass is more complicated. For $\beta > 0$, this indicates that if $\Pi(h) \leq 0$ over the domain, then (11) yields a positive growth rate, namely condensation. In critical case where the sign of $\Pi(h)$ is changeable, transition between condensation and evaporation can occur.

Fig. 1 shows dynamics of (1) starting from a typical initial profile, $h_0(x) = 0.2 - 0.03 \sin(6\pi x/L) + 0.01 \sin(4\pi x/L)$, on the periodic domain $0 \leq x \leq L$ with $L = 20$. Initially, the long-wave instability due to the disjoining pressure $\Pi(h)$ yields coalescence of the film into three isolated droplets connected by a thin layer $h \sim \epsilon$. Due to the competition between the evaporation effects and the disjoining pressure, the right and the left droplets collapse to the thin layer sequentially, followed by the slow condensation of the center droplet. This drop-wise condensation mode then transitions to film-wise condensation as the condensing droplet reaches the size of the domain. Motivated by the interesting combination of dewetting and evaporative/condensing effects shown in Fig. 1, we will explore how such dynamics transitions occur, starting with examining the instability of both spatially uniform and non-uniform equilibrium solutions.

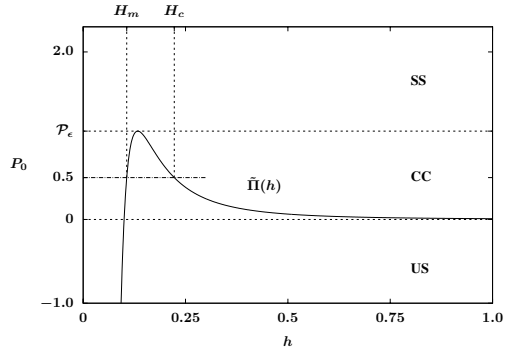

FIG. 3. A plot of $\tilde{\Pi}(h)$ given by (5) with $\epsilon = 0.1$ with the dashed lines representing the thresholds of parameter $P_0$ which determines the multiplicity of spatially-uniform solutions in the system. The thresholds separate the under-saturated (US), critical case (CC) and super-saturated (SS) regimes.

## III. STABILITY OF SPATIALLY UNIFORM FILMS

We begin by examining the dynamics for solutions that are perturbations of flat fluid layers. Since mass is not conserved, the mean height, $\bar{h}(t)$, can evolve with time and there is a need to cover a range of behaviors spanning from evaporating down to uniform adsorbed layers to flooding under condensation.

We follow the approach in [24, 43] and perturb the uniform film by an infinitesimal Fourier mode disturbance

$$h(x,t) \sim \bar{h}(t) + \delta e^{\mathrm{i}2k\pi x/L} e^{\sigma(t)}, \tag{12}$$

where $k$ is the wavenumber and $\sigma(t)$ describes the growth of the perturbation starting from the initial amplitude $\delta \ll 1$, with $\sigma(0) = 0$. Expanding PDE (1) about $h = \bar{h}$ then gives the $O(1)$ and $O(\delta)$ equations:

$$O(1): \qquad \frac{d\bar{h}}{dt} = -\beta \tilde{J}(\bar{h}) \quad \text{where } \tilde{J}(\bar{h}) = \frac{\Pi(\bar{h})}{\bar{h} + K_0}, \tag{13a}$$

$$O(\delta): \qquad \frac{d\sigma}{dt} = -\left[ \left( \frac{2k\pi}{L} \right)^2 \bar{h}^3 + \frac{\beta}{\bar{h} + K_0} \right] \left[ \Pi'(\bar{h}) + \left( \frac{2k\pi}{L} \right)^2 \right] + \frac{\beta}{\bar{h} + K_0} \tilde{J}(\bar{h}), \tag{13b}$$

For $\beta \neq 0$, the equilibria of (13a) are the spatially uniform steady states of (1), namely the heights yielding zero pressure, $\Pi(\bar{h}) = 0$. As shown in Fig. 3, the number of spatially uniform steady states depends on the value of the parameter $P_0$ relative to the critical pressure $\mathcal{P}_\epsilon$ set by the disjoining pressure, from the unique maximum at $h_{\mathrm{peak}} = \frac{4}{3}\epsilon$ where

$$\Pi(h_{\mathrm{peak}}) = \mathcal{P}_\epsilon - P_0 \quad \text{and} \quad \mathcal{P}_\epsilon = \frac{27}{256}\epsilon^{-1} > 0. \tag{14}$$

In particular, there are three regimes:

1. Undersaturated vapor (US): If $P_0 < 0$, there is one uniform steady state with $H_m < h_{\text{peak}}$,

2. Critical case (CC): If $0 < P_0 \leq \mathcal{P}_\epsilon$, there are two uniform steady states with $H_m \leq h_{\text{peak}} \leq H_c$,

3. Supersaturated vapor (SS): If $P_0 > \mathcal{P}_\epsilon$, there is no uniform steady state.

These cases for the range of $P_0$ are of central importance for both the stability analysis and dynamics of this problem. For $P_0 < \mathcal{P}_\epsilon$ in the critical and undersaturated cases, the spatially uniform steady state $H_m$ defines a minimum thickness adsorbed layer,

$$H_m = \epsilon + P_0 \epsilon^2 + 4P_0^2 \epsilon^3 + O(\epsilon^4) \qquad \epsilon \to 0, \tag{15}$$

while in the critical case, for $0 < P_0 < \mathcal{P}_\epsilon$, another constant steady state $H_c$ also exists,

$$H_c = P_0^{-1/3} \epsilon^{2/3} - \tfrac{1}{3}\epsilon - \tfrac{2}{9} P_0^{1/3} \epsilon^{4/3} + O(\epsilon^{5/3}). \tag{16}$$

While $H_c$ depends on $P_0$ in the leading order, $H_m$ has a weaker dependence on $P_0$ with a saddle-node bifurcation occurring at $P_0 = \mathcal{P}_\epsilon$. Similar results on flat film steady states are available in [36, 43] where other forms for the disjoining pressure were used. For the rest of this article, we set the system parameters $\epsilon = K_0 = 0.1$, yielding $\mathcal{P}_\epsilon \approx 1.05$.

The equilibria $H_m, H_c$ are fixed points of (13a) and then (13b) reduces to the dispersion relation for the growth rate of perturbations $\lambda$ from standard linear stability analysis,

$$\sigma(t) = \lambda t, \qquad \lambda(\bar{h}) = -\left[ \left( \frac{2k\pi}{L} \right)^2 \bar{h}^3 + \frac{\beta}{\bar{h} + K_0} \right] \left[ \Pi'(\bar{h}) + \left( \frac{2k\pi}{L} \right)^2 \right]. \tag{17}$$

For $\bar{h} = H_m$, $\Pi'(\bar{h}) > 0$ to yield that $\lambda \leq 0$ so $H_m$ is stable with respect to all perturbations. For $\bar{h} = H_c$, $\Pi'(\bar{h}) < 0$ and the thickness $H_c$ is long-wave unstable with respect to perturbations with $0 \leq k < k_c$ for a critical wavenumber $k_c = L|\Pi'(H_c)|^{1/2}/(2\pi)$.

Linear stability analysis of films with other thicknesses has been previously carried out using the frozen time (or quasi-steady $\bar{h}$) approach [14, 36, 46], but this is limited to only short-time behaviors. We will consider the evolution more globally by using the full dynamics of (13a, 13b) for the mean film thickness and amplitude of spatial perturbations.

The loss or gain of mass and evolution of the mean film thickness $\bar{h}(t)$ follows directly from (13a) and the ranges of $\bar{h}$ for which $\Pi(h)$ is positive or negative, see Fig. 4. For the critical and undersaturated cases, the local linear stability/instability of $H_m$ and $H_c$ extend to show those states to be dividing lines between regimes of condensation and evaporation for uniform films. For the supersaturated case, $\Pi > 0$ for all $\bar{h} > 0$ and hence all films will exhibit condensation.

We now turn to (13b) to investigate the possibility for development of spatial patterns on the film surface. Fig. 5(left) shows the growth rate $d\sigma/dt$ as a function of $\bar{h}$ for several values of $P_0$ at a given perturbation wavenumber, $k = 6$. We observe that there exists a threshold pressure $P_0 = P_c(k)$ for the growth of spatial perturbations. For $P_0 < P_c$, films of thickness $\bar{h}$ are linearly unstable to perturbations over a band of average film thicknesses

$$\bar{h}_-(P_0, k) < \bar{h} < \bar{h}_+(P_0, k), \tag{18}$$

for which $d\sigma/dt > 0$ (see Fig. 5(right)). To see an illustration of the impact of this instability, we consider an example with $P_0 = 0.5$ and initial condition $h_0(x) = 0.2 + 10^{-4} \cos(12\pi x/L)$ in (1). Here, the initial mean thickness is $\bar{h}_0 = 0.2$ which lies in the unstable range $\bar{h}_- < \bar{h} < \bar{h}_+$ for $k = 6$ (see Fig. 5(right)). The value of $\bar{h}_0$ is also less than $H_c(P_0) \approx 0.222$, so by (13a) we expect evaporation to occur. To compare the linearized model against direct simulations of (1), we define the average film thickness,

$$\langle h \rangle = \frac{1}{L} \int_0^L h \, dx. \tag{19}$$

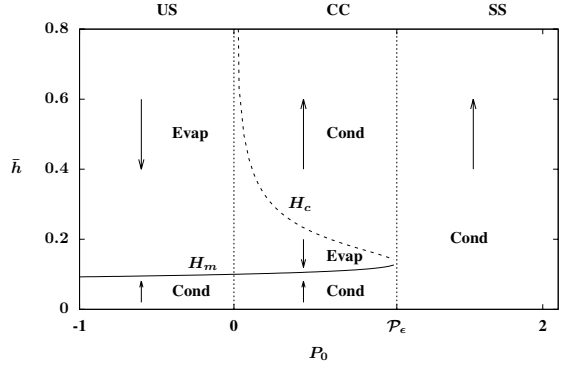

FIG. 4. The bifurcation diagram of the uniform steady states $H_c$ and $H_m$ as functions of $P_0$. Arrows indicate the dynamics for the mean thickness $\bar{h}(t)$ in each of the regimes, US/CC/SS: condensation with $\bar{h}(t)$ increasing or evaporation with $\bar{h}(t)$ decreasing, and $H_m$ being stable while $H_c$ is an unstable steady state.

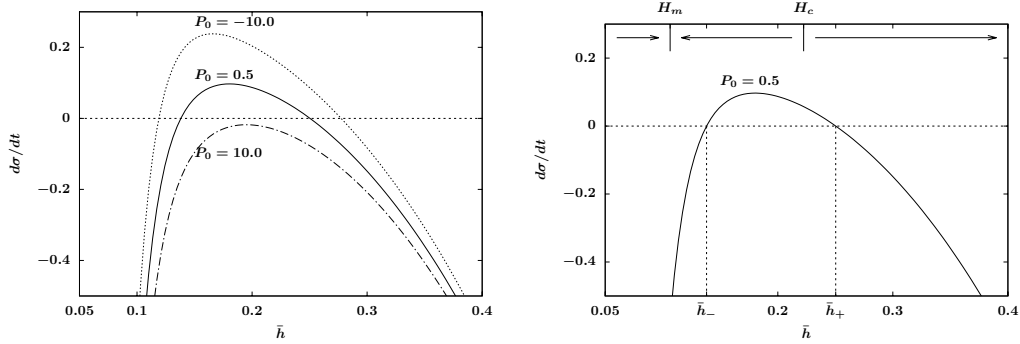

FIG. 5. (Left) A plot of $d\sigma/dt$ vs $\bar{h}$ at several values of $P_0$ (with $k = 6$, $\beta = 0.001$ and $L = 20$ fixed). Here $P_c \approx 8.455$ is a threshold for growth of spatial instabilities for $P_0 < P_c$. (Right) Critical film heights $\bar{h}_-, \bar{h}_+$ marking the range where spatial perturbations grow for $P_0 = 0.5$. This range of heights is different than the ranges for evaporation or condensation of uniform films set by the equilibria $H_m$, $H_c$.

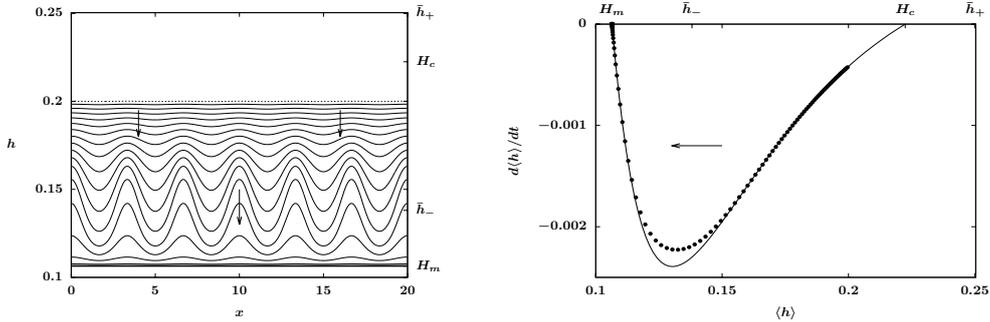

FIG. 6. (Left) Evaporation with transient pattern formation starting from the initial data $h_0(x) = 0.2 + 10^{-4}\cos(12\pi x/L)$ with the system parameters $P_0 = 0.5$, $\beta = 0.001$ and $L = 20$. (Right) The dots represent the effective rate of evaporation $d\langle h \rangle/dt$ plotted against the decreasing average film thickness in the PDE simulation, compared against the prediction $-\beta \tilde{J}(\langle h \rangle)$ from (13a) in solid line.

The simulation in Fig. 6(left) shows transient pattern formation with the perturbation amplitude growing while $\bar{h}(t)$ evolves downward through the unstable range. As $\bar{h}(t)$ crosses $\bar{h}_-$ the perturbations decay as the solution eventually converges to the spatially uniform steady state $H_m$. In Fig. 6(right) we compare the prediction for the evaporation rate from (13a) against the effective rate exhibited by the average thickness, $d\langle h\rangle/dt \approx -\beta\tilde{J}(\langle h\rangle)$. We observe that for this example the assumptions in (13a, 13b) hold well over most of the dynamics except for an intermediate period. When $\langle h\rangle \approx 0.13$ the spatial perturbations became sufficiently large to have nonlinear effects making deviations between $\bar{h}$ and $\langle h\rangle$ noticeable (with the nonlinear spatial perturbations having non-zero means).

To understand when the influences of transients will be significant, we examine the behavior described by (13b). Since we know that $\bar{h}$ evolves monotonely between the equilibria $H_c, H_m$, it is helpful to combine (13a, 13b) to form the associated ODE for the evolution of the spatial perturbation amplitude with respect to changes in the mean thickness,

$$\frac{d\sigma}{d\bar{h}} = \frac{1}{\tilde{J}}\left[\frac{1}{\beta}\left(\frac{2k\pi}{L}\right)^2\bar{h}^3 + \frac{1}{\bar{h}+K_0}\right]\left[\Pi'(\bar{h}) + \left(\frac{2k\pi}{L}\right)^2\right] - \frac{1}{\bar{h}+K_0}. \tag{20}$$

In the limit of weak evaporation/condensation effects, this takes the form

$$\frac{d\sigma}{d\bar{h}} \sim \frac{\bar{h}+K_0}{\beta\Pi(h)}\left(\frac{2k\pi}{L}\right)^2\bar{h}^3\left[\Pi'(\bar{h}) + \left(\frac{2k\pi}{L}\right)^2\right] = O\left(\frac{1}{\beta}\right) \quad \text{for } \beta \to 0, \tag{21}$$

and hence we can conclude that for $\beta \to 0$ spatial perturbations can display large transient growth relative to given changes of the mean film height. Since the factor in square brackets in (21) is the same as in (17), we see that these transient instabilities share the same critical wavenumber $k_c$ for long-wave instability. In general the limit of weak phase change effects ($\beta \to 0$) is a singular perturbation: for finite times (1) with $\beta \to 0$ approaches the nonvolatile behavior of (2), but for sufficiently long times ($t \to \infty$) change in mass will create significant differences in behaviors.

In contrast, when $\beta$ is large,

$$\frac{d\sigma}{d\bar{h}} \sim \frac{1}{\Pi(\bar{h})}\left[\Pi'(\bar{h}) + \left(\frac{2k\pi}{L}\right)^2\right] - \frac{1}{\bar{h}+K_0} = O(1) \quad \text{for } \beta \to \infty, \tag{22}$$

and there is no separation of time-scales between the evolution of $\bar{h}(t)$ and $\sigma(t)$. This means that with strong evaporation, perturbations would not have a chance to appreciably grow in amplitude before $\bar{h}(t)$ has moved out of the unstable range and hence rapidly evaporating films would always stay close to uniform in height.

Fig. 7 gives the dispersion relation (13b) for several values of $\bar{h}$. We observe that for a range of $\bar{h}$ values the most unstable wavenumber is positive, $k^* > 0$. This means that patterns can develop for those thicknesses (see the $\bar{h} = 0.2$ and $\bar{h} = 0.3$ curves). Below some threshold for $\bar{h}$, the most unstable wavenumber is attained at $k^* = 0$, indicating that any spatial perturbations will decay relative to the evolution of the uniform film $\bar{h}(t)$ (see the $\bar{h} = 0.127$ and $\bar{h} = 0.133$ curves). For zero wavenumber these perturbations are constant in space, $\bar{h} \to \bar{h} + \delta e^\sigma$, this effectively reduces to a sensitivity analysis for (13a) with respect to the evolving base state $\bar{h}(t)$: $d\sigma/dt = -\beta\tilde{J}'(\bar{h})$. Similar analysis for the stability of dewetting evaporating thin films has also been done in [48]. The transitions between these $\bar{h}$-dependent forms of dispersion relations occurs at $k^* = 0$ and appears similar to the type-II and type-III instabilities described in [16]. The significant qualitative changes in the dispersion relation as $\bar{h}(t)$ evolves shows that the frozen-time approximation cannot be expected to give good predictions for the dynamics in the critical pressure regime.

In Fig. 8 we demonstrate that under the same system parameters, starting with a thicker film ($\langle h_0 \rangle = 0.3 > H_c$) yields condensation. With $P_0 = 0.5$ in the critical range, the numerical simulation for (1) starting from the weakly perturbed film (initial profile given in the figure caption) shows monotone growth of the mass. The apparent most unstable wavenumber $k^*$ is numerically tracked by using the Fourier transform of the time derivative of the PDE solution $\partial h/\partial t$ and is plotted against the average film thickness. Starting from $k^* = 6$, the evolution going down to $k^* = 0$ is shown overlaid on the contour lines of the dispersion relation (13b) and follows the linear prediction well. This is expected, since while there may be some transient growth of spatial modes, as condensation continues and $\bar{h}(t)$ increases further, the perturbations are relatively small in amplitude and will also eventually decay out. So transient pattern formation also can occur in condensation but it is not as dramatic as in evaporation.

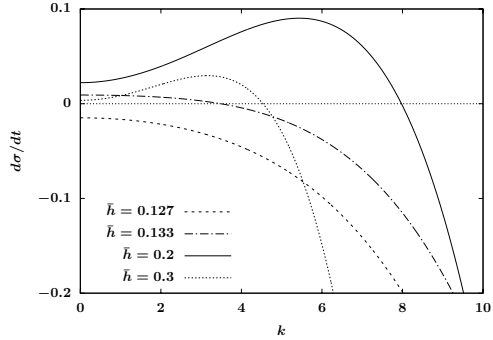

FIG. 7. The dispersion relation plot of $d\sigma/dt$ against the wave number $k$ at the values $\bar{h} = 0.127, 0.133, 0.2, 0.3$. The parameter $P_0 = 0.5$ is in the critical case range $0 < P_0 < \mathcal{P}_c$ and the other parameters are identical to those in Fig. 5.

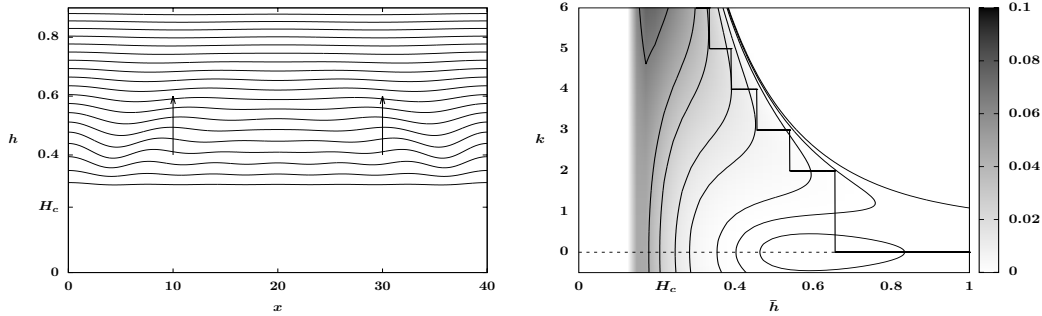

FIG. 8. (Left) Condensation dynamics with transient pattern formation starting from the initial data $h_0(x) = 0.3 + 0.001 \sum_{k=1}^{6} \cos(2k\pi x/L)$ with $L = 40$ and other parameters identical to Fig. 6 (Right) Contour plot of the growth rate $d\sigma/dt$ with the thick line representing the computed most unstable wavenumber $k^*$ as the condensation occurs.

The long-time behavior for condensation is unbounded growth of the film thickness (sometimes called "flooding"), where $\bar{h} \to \infty$. In this limit, equation (13a) requires $P_0 > 0$ and hence this behavior only occurs in the critical and super-saturated cases, with the flux reducing to $\tilde{J} \sim -P_0/\bar{h}$. For $\bar{h} \to \infty$, (13b) reduces to $d\sigma/dt \sim -\bar{h}^3(2k\pi/L)^4$, namely for all $\beta$ (and all $P_0 > 0$), spatial perturbations will decay due to surface tension and flooding will always be manifested in terms of flat films following the leading order behavior

$$\frac{d\bar{h}}{dt} \sim \frac{\beta P_0}{\bar{h}} \qquad \Longrightarrow \qquad \bar{h}(t) \sim \sqrt{2\beta P_0 t} \quad \text{for } t \to \infty. \tag{23}$$

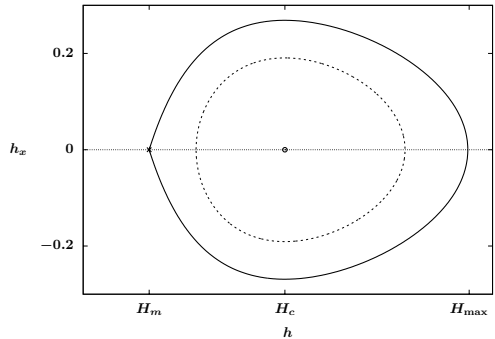

FIG. 9. The phase plane for the second-order steady state solutions satisfying (24) in the critical case with $0 < P_0 < \mathcal{P}_\epsilon$. The homoclinic orbit in solid line corresponds to the droplet solution in Fig. 10 (left) and the dashed line represents the periodic solution in Fig. 10 (right).

## IV. NON-UNIFORM STEADY STATES

Recalling the energy (9), all steady states of (1) have zero pressure, and the periodic non-uniform steady states satisfy the boundary value problem on $0 \le x \le L$,

$$\frac{d^2 h}{dx^2} - \Pi(h) = 0, \qquad h(0) = h(L), \qquad h'(0) = h'(L). \tag{24}$$

For the undersaturated case with $P_0 < 0$ and the supersaturated case with $P_0 > \mathcal{P}_\epsilon$, equation (24) does not have any non-uniform solutions. In this section we will obtain non-trivial periodic solutions to (24) that exist only for the critical case with $0 < P_0 < \mathcal{P}_\epsilon$.

Steady states of the mass-conserving thin film equation (2) satisfy the same form of second-order differential equation problem (24). For the conservative problem with $\beta = 0$, $P_0$ in (5) is a free parameter that parametrizes the continuous family of coexisting steady states. For the non-conservative model (1) with $\beta > 0$, $P_0$ is an imposed system parameter that determines a unique non-uniform steady state on a given domain.

With different forms of $\tilde{\Pi}(h)$ in (5), the steady states and stability of conservative thin film equations have been the focus of many studies. In [11] Bertozzi et al. considered the steady states using the regularized disjoining pressure $\tilde{\Pi}(h) = Ah^{-3}(1 - \epsilon/h)$. Laugesen and Pugh studied the properties and energy levels of positive steady states for a generalized conservative thin film equation with $\tilde{\Pi}(h) = h^{-\alpha}$ [27, 28], and provided linear stability results for a non-conservative thin film model with a linear fourth-order term [26]. In [28] Laugesen and Pugh showed that the linear stability of even steady states that satisfy periodic boundary conditions on $(0, L)$ is equivalent to the linear stability of the steady states with respect to the Neumann (or 'no flux') boundary conditions over the half domain $0 \le x \le L/2$.

The phase plane of $(h, h_x)$ is plotted in Fig. 9 where the equilibrium $H_m(P_0)$ is a saddle point, and $H_c(P_0)$ is a center point. Picking a value of the minimum height $h_{\min}$ in the range $H_m \le h_{\min} \le H_c$ determines a contour in the phase plane describing a periodic steady state solution. For each value of $h_{\min}$, the corresponding maximum height $h_{\max}$ comes from $U(h_{\max}) = U(h_{\min})$. In particular, if $h_{\min} = H_m$, we have a homoclinic orbit through the saddle point $H_m$ which corresponds to a solitary droplet solution shown in Fig. 10 (left). The maximum height $H_{\max}$ of the droplet is then given by the root of $U(H_{\max}) = U(h_{\min})$, which for $\epsilon \to 0$ can be written as

$$H_{\max} = \frac{1}{6P_0} + \epsilon + O(\epsilon^2). \tag{25}$$

For $H_m < h_{\min} < H_c$, we have periodic steady state solutions with a typical plot presented in Fig. 10 (right). The length of the period, $\ell(h_{\min})$, of the periodic steady state $h(x)$ can be obtained

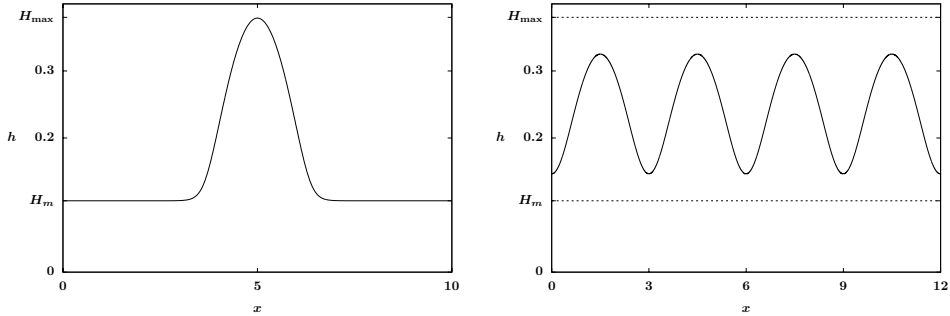

FIG. 10. (Left) A typical steady-state droplet solution and (right) a periodic solution satisfying the ODE (24).

by

$$\ell(h_{\min}) = \int_0^{\ell(h_{\min})} dx = 2 \int_{h_{\min}}^{h_{\max}} \frac{1}{\sqrt{2U(s) - 2U(h_{\min})}} \, ds. \tag{26}$$

Small amplitude periodic steady state solutions $h$ bifurcate from the spatially uniform steady state $H_c$ [48]. Using linear stability at $H_c$ it can be shown [11, 41] that the minimum period of oscillations is given by

$$\ell_s = \frac{2\pi}{\sqrt{-\Pi'(H_c)}}. \tag{27}$$

Equation (26) defines a monotone function on $h_{\min}$ satisfying

$$\ell(h_{\min} \to H_c) = \ell_s, \qquad \ell(h_{\min} \to H_m) \to \infty.$$

We denote primary periodic steady state solutions with a single maximum on domain $0 \le x \le L$ as $h = H_s(x)$. Fig. 11 shows that the average thickness, $\langle H_s \rangle$, parameterized by $L$, bifurcates from $H_c$ at $L = \ell_s$. Since $H_s(x)$ approaches a finite-mass solitary droplet, we have $\lim_{L \to \infty} \langle H_s \rangle = H_m$. Higher order families of non-uniform steady states $H_{s,k}(x)$ with $k$ periods bifurcating from $H_c$ at $L = k\ell_s$ for $k = 2, 3, \cdots$ are also shown in Fig. 11. Multiple periodic non-uniform steady states can coexist on a given domain with $L \ge 2\ell_s$.

The value of the system parameter $P_0$ in also important in selecting between multiple coexisting periodic steady states $H_s(x)$. With a fixed domain size $L = \ell = 3.05$, a bifurcation diagram of the average film thickness of steady states satisfying (24) with respect to $P_0$ is presented in Fig. 12 (left). It shows that in addition to the constant steady states $h \equiv H_c$ and $h \equiv H_m$, a family of non-uniform steady state solutions $H_s(x)$ exist for $P^\circ \le P_0 \le P^\bullet$ bifurcating from the uniform steady state $H_c$ at $P_0 = P^\circ$ and $P_0 = P^\bullet$. These two bifurcation points can be obtained by setting $\ell_s = L$ and solving (27) for the $P_0$ values. Moreover, there exists a critical $P^* = P^*(\ell)$ such that for $P^* < P_0 < P^\bullet$ the average thickness $\langle H_s \rangle$ of $H_s(x)$ is monotonically decreasing with respect to the parameter value $P_0$, while for $P^\circ < P_0 < P^*$ we have $\langle H_s \rangle$ increasing with $P_0$. The maximum value $\langle H^* \rangle$ of the average thickness is attained at $P_0 = P^*$, and the average thickness of small amplitude steady state solutions are given by $\langle H^\circ \rangle$ and $\langle H^\bullet \rangle$ at $P_0 = P^\circ$ and $P_0 = P^\bullet$, respectively.

The above bifurcation diagram also characterizes the coexistence of steady states with given length of period and average film thickness. Specifically, it indicates that for $\langle H^\circ \rangle < \langle h \rangle < \langle H^* \rangle$, two non-uniform steady states and the uniform state $H_c$ coexist. An example of coexisting steady states with $\langle h \rangle = 0.24$ are shown in Fig. 12 (right), which corresponds to three marked dots in Fig. 12 (left). This is consistent with the results obtained in [27] on the existence and uniqueness of steady state solutions of the mass-conserving thin film model with specific domain size and average thickness.

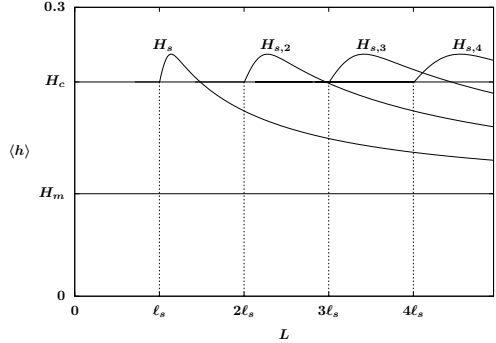

FIG. 11. Bifurcation diagram for $\langle h \rangle$ parametrized by the domain size $L$ with the parameter $P_0 = 0.5$.

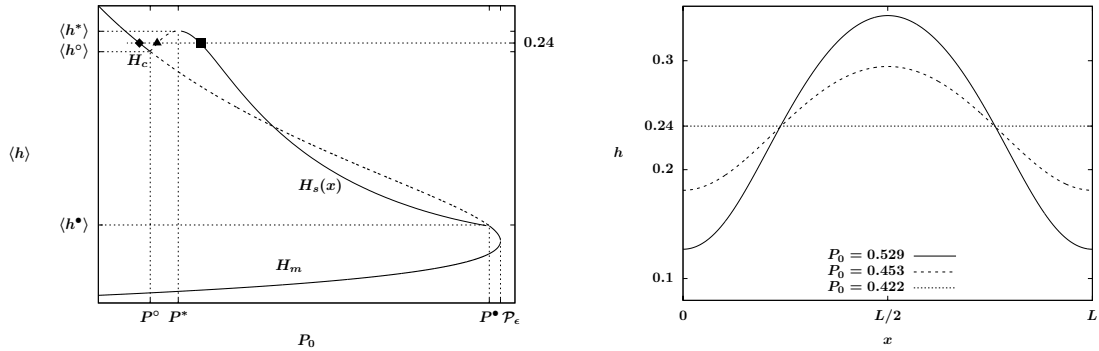

FIG. 12. A bifurcation diagram of steady states to (24) with the domain size $L = 3.05$ showing that a family of non-uniform steady states coexist with the constant steady states $H_m$ and $H_c$ in a specific range of $P_0$. The solid lines represent the stable steady states and the dashed lines correspond to the unstable ones. (Right) coexisting equilibria of (24) with the average film thickness $\langle h \rangle = 0.24$ and the domain size $L = 3.05$.

## V. STABILITY ANALYSIS OF NON-UNIFORM STEADY STATES

In the previous section we have studied the steady states, and in this section we focus on their linear stability and transient behavior. We consider a positive $\ell$-periodic steady state $H_s(x)$ over the domain $0 \leq x \leq L$, and perturb it by setting $h(x,t) = H_s(x) + \delta \Psi(x) e^{\lambda t}$, where $\delta \ll 1$ and $\Psi(x)$ is also $\ell$-periodic. Since $H_s(x)$ satisfies the ODE (24), we linearize the equation (1) around the steady state $H_s(x)$ and obtain the $O(\delta)$ equation

$$\lambda \Psi = \mathscr{L} \Psi \tag{28}$$

where the linear operator $\mathscr{L}$ is

$$\mathscr{L} \Psi \equiv \left[ -\frac{\beta}{H_s + K_0} + \frac{d}{dx} \left( H_s^3 \frac{d}{dx} \right) \right] \left( \Pi'(H_s)\Psi - \frac{d^2\Psi}{dx^2} \right). \tag{29}$$

We rewrite this fourth-order operator as the product of two second-order operators,

$$\mathscr{L} \Psi = \mathscr{Q}\mathscr{P}\Psi, \tag{30}$$

where $\mathscr{Q}$ is an operator that includes the evaporative flux and the mobility,

$$\mathscr{Q}w \equiv -\frac{\beta w}{H_s + K_0} + \frac{d}{dx} \left( H_s^3 \frac{dw}{dx} \right), \tag{31}$$

and $\mathcal{P}$ is the linearized pressure operator,

$$\mathcal{P}v \equiv \Pi'(H_s)v - \frac{d^2v}{dx^2}. \tag{32}$$

The operator $\mathscr{L}$ is not self-adjoint with respect to the standard $L^2$ inner product; the adjoint operator $\mathscr{L}^\dagger$ is given by

$$\mathscr{L}^\dagger \Phi \equiv \mathcal{P}\mathcal{Q}\Phi. \tag{33}$$

If there are any positive eigenvalues $\lambda$ to the problem (28), then the steady state $H_s(x)$ is unstable. In section V A we will study this eigenvalue problem for the conservative model with $\beta = 0$ where $\mathscr{L}$ in (29) reduces to a simpler operator. Then through an operator expansion in the limit $\beta \to 0$ we then obtain the stability of the non-uniform steady states in section V B. For simplicity in this section we only focus on the case where a one period solution fits in the domain, that is, $L = \ell$, but this framework can be easily extended to solutions with multiple periods.

## A. Linear stability of the conservative PDE with $\beta = 0$

The linear stability analysis of the steady states of the conservative PDE (2) with $\beta = 0$ is important to the understanding of the stability of equilibria of the model (1) with $\beta \ll 1$, so here we begin with a brief overview of the stability of the steady states of model (2) [11]. We will derive a critical domain size $L = \ell^*$ where the stability of the steady states change for $\beta = 0$.

The linear stability of these non-uniform steady states $H_s(x)$ of (2) can be obtained by solving the eigenproblem (28) which by using $\beta = 0$ reduces to

$$\mathscr{L}_0 \Psi = \lambda \Psi, \qquad \text{where } \mathscr{L}_0 \Psi \equiv \frac{d}{dx}\left[H_s^3 \frac{d}{dx}\left(\mathcal{P}\Psi\right)\right]. \tag{34}$$

We numerically solve this eigenproblem for the steady states presented in Fig. 12 (left) and show that for the range $P^* < P_0 < P^\bullet$ the family of non-uniform steady states is stable. It coexists with the branch of spatially-uniform unstable steady state $h \equiv H_c$, and becomes unstable for $P^\circ < P_0 < P^*$. While the stability of the uniform steady states $h \equiv H_m$ is independent of $P_0$, the other steady state $h \equiv H_c$ is only stable in the ranges $P^\bullet < P_0 < \mathcal{P}_\epsilon$ and $P_0 < P^\circ$, and is unstable for $P^\circ < P_0 < P^\bullet$, where the non-uniform steady state exists. This difference from the stability of $H_c$ described in section III is due to the conservation of mass condition for $\beta = 0$ (eliminating the $k = 0$ mode) and periodic boundary conditions on a finite domain selecting discrete modes.

Numerical evidence suggests that the critical value $P^*$ is monotonically decreasing with respect to $\ell$ for $\ell > \ell_s$. Therefore given a fixed value of the parameter $P_0$, there exists a critical period $\ell^*$ such that $P^*(\ell^*) = P_0$. Then for any period $\ell$ in the range of $\ell_s < \ell < \ell^*$, we have $P^*(\ell) > P^*(\ell^*) = P_0$. This indicates that the corresponding steady state $H_s(x)$ with the period $\ell$ and average film thickness $\langle H_s \rangle$ is unstable with a positive eigenvalue $\lambda$ to the eigenproblem (34). There coexists another steady state which is stable with $P_0 > P^*(\ell)$ with the same period $\ell$ and average film thickness $\langle H_s \rangle$. On the other hand, for $\ell > \ell^*$ with $P^*(\ell) < P^*(\ell^*) = P_0$, there exists a stable steady state $H_s(x)$, and the corresponding steady state with $P_0 < P^*(\ell)$ is unstable.

For the parameter $P_0 = 0.5$, we numerically calculate the corresponding critical period as $\ell^* = 3.005$. With this parameter choice in Fig. 13 we plot the two largest eigenvalues of the eigenproblem (34) for the non-uniform steady state $H_s(x)$ over a range of periods $\ell$. In addition to the $\lambda = 0$ translational mode, for $\ell_s < \ell < \ell^*$ we have a positive eigenvalue making $H_s$ unstable while for $\ell > \ell^*$ $H_s$ is stable.

The relationship between the period of a given steady state and the critical length of period $\ell^*$ is important to the stability of the PDE (1) for both the cases $\beta = 0$ and $\beta \neq 0$. In the following section we focus on the influence of weak non-conservative effects and for $\beta > 0$ we will see the same change in stability at $\ell = \ell^*$.

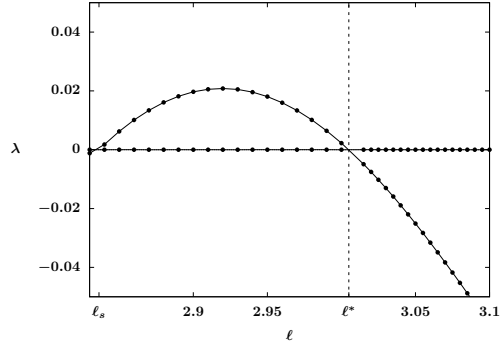

FIG. 13. The dependence of dominant eigenvalues of $H_s(x)$ on the period $\ell$ for the eigenproblem (34) with $P_0 = 0.5$, where the critical period is given by $\ell^* = 3.005$.

## B. Non-uniform steady states: linear stability with $\beta \to 0$

For non-uniform steady states $H_s(x)$ with $\beta \neq 0$ we first derive some properties for the eigenfunctions of $\mathscr{L}$ (28), and then characterize its dominant eigenmodes using an asymptotic operator expansion as $\beta \to 0$. By taking the $L^2$ inner product of $\mathscr{L}\Psi$ and $\mathcal{P}\Psi$. Integration by parts and using $\beta > 0$ then lead to

$$\langle \mathscr{L}\Psi, \mathcal{P}\Psi \rangle = -\beta \int_0^L \frac{(\mathcal{P}\Psi)^2}{H_s + K_0} \, dx - \int_0^L H_s^3 \left[ \frac{d}{dx} \left( \mathcal{P}\Psi \right) \right]^2 \, dx \leq 0. \tag{35}$$

Applying (28) to the inner product and using integration by parts, one gets

$$\langle \mathscr{L}\Psi, \mathcal{P}\Psi \rangle = \lambda \langle \Psi, \mathcal{P}\Psi \rangle = \lambda \int_0^L \Pi'(H_s)\Psi^2 + \Psi_x^2 \, dx. \tag{36}$$

Combining (35) and (36) we conclude that the eigenfunctions associated with positive eigenvalues $\lambda > 0$ satisfy

$$\int_0^L \Pi'(H_s)\Psi^2 + \Psi_x^2 \, dx \leq 0.$$

Moreover, if $\lambda = 0$, then the two integrals in (35) indicate that the eigenfunction $\Psi$ associated with zero eigenvalue is determined by

$$\mathcal{P}\Psi \equiv \Pi'(H_s)\Psi - \frac{d^2\Psi}{dx^2} = 0. \tag{37}$$

The solution of (37) leads to a translational eigenfunction associated with zero eigenvalue, $\Psi^T = \frac{dH_s}{dx}$.

For the special case with $\beta = 0$ when $P_0$ can take a continuous range of values, there is an additional eigenfunction $\Psi_0^P = \frac{dH_s}{dP_0}\bigg|_{\beta=0}$ in the nullspace of the linearized operator $\mathscr{L}$. Taking derivative of the ODE (24) with respect to $P_0$ one gets $\mathcal{P}\Psi_0^P = 1$, which then leads to $\mathscr{L}\Psi_0^P = \mathcal{Q}(1) = 0$.

To understand the dependence of the spectrum of $\mathscr{L}$ on the parameter $\beta$, we consider the following asymptotic expansion as $\beta \to 0$,

$$\Psi = \Psi_0 + \beta\Psi_1 + \cdots, \qquad \lambda = \lambda_0 + \beta\lambda_1 + \cdots. \tag{38}$$

Substituting these expansions into the eigenproblem (28) leads to the $O(1)$ problem

$$\mathscr{L}_0 \Psi_0 = \lambda_0 \Psi_0, \tag{39}$$

which is equivalent to the eigenproblem (34) for the conservative thin film equation (2) without any mass conservation constraint. It has been shown in [10, 55] that the linear operator $\mathscr{L}_0$ is self-adjoint with respect to a weighted $H^{-1}$ norm. Therefore from spectral theory, for a compact domain the associated spectrum of the leading order eigenvalue problem (39) is real and discrete. The $O(\beta)$ problem can be written as

$$\mathscr{L}_0 \Psi_1 - \lambda_0 \Psi_1 = \frac{\mathcal{P}\Psi_0}{H_s + K_0} + \lambda_1 \Psi_0. \tag{40}$$

Integrating (39) over the domain gives

$$\lambda_0 \int_0^L \Psi_0 \; dx = 0, \tag{41}$$

from which we see that any nonzero eigenvalue of (39) leads to a zero-mean eigenfunction, and if $\Psi_0$ has a non-zero mean then the corresponding $\lambda_0 = 0$. The translational mode $\Psi_0^T = \frac{dH_s}{dx}$ has a zero mean due to the periodic boundary conditions, and has $\lambda_0^T = 0$. The pressure mode $\Psi_0^P = \frac{dH_s}{dP_0}\big|_{\beta=0}$ has $\lambda_0^P = 0$ and has a non-zero mean. We denote the largest nonzero eigenvalue as $\lambda_0^V$ whose associated eigenmode $\Psi_0^V$ has zero mean and leads to growth or decay of spatial variations from the mean of the solution $h(x,t)$.

For the translational eigenfunction $\Psi^T$ the expansion (38) is trivial since this branch of zero eigenvalue is independent of the value of $\beta$ and exists for any periodic solutions.

For the second eigenfunction $\Psi^P$ we can integrate (40) over the domain and apply the periodic boundary conditions to obtain

$$\lambda_1^P = -\frac{1}{\langle \Psi_0^P, 1\rangle} \left\langle \frac{\mathcal{P}\Psi_0^P}{H_s + K_0}, 1 \right\rangle.$$

For the $\Psi^V$ mode, integrating (40) over $0 \leq x \leq L$ and applying (41) then yields

$$\int_0^L \Psi_1^V \; dx = -\frac{1}{\lambda_0^V} \int_0^L \frac{\mathcal{P}\Psi_0^V}{H_s + K_0} \; dx. \tag{42}$$

This is not guaranteed to be zero, therefore for $\beta \neq 0$ we expect to have eigenfunctions of (28) with nonzero means which account for the loss or gain of mass in the dynamics. Moreover, in this case $\lambda_1^V$ can be determined by applying the solvability condition to (40). Note that the adjoint operator $\mathscr{L}_0^\dagger$ of the leading order operator $\mathscr{L}_0$ takes the form $\mathscr{L}_0^\dagger \Phi_0^V \equiv (\mathcal{P}\mathcal{Q}|_{\beta=0}) \Phi_0^V$, where the eigenfunction of the adjoint operator associated with the leading order eigenvalue $\lambda_0^V$ is denoted as $\Phi_0^V$. Taking the inner product of $\Phi_0^V$ with the right-hand-side of (40) gives

$$\lambda_1^V = -\frac{1}{\langle \Psi_0^V, \Phi_0^V\rangle} \left\langle \frac{\mathcal{P}\Psi_0^V}{H_s + K_0}, \Phi_0^V \right\rangle. \tag{43}$$

We now investigate the linear stability of the non-uniform steady state $H_s(x)$ where the domain size $L$ is either in the range of $L > \ell^*$ or $\ell_s < L < \ell^*$. Specifically, we are interested in the bifurcation of the eigenproblem (28) parametrized by $\beta$ in the limit of $\beta \to 0$. Recall from section V A that the parameter $P_0 = 0.5$ corresponds to the critical $\ell^* = 3.005$. We will use this parameter value for the following discussion. To demonstrate that $\ell^* = 3.005$ is indeed a critical value for the non-uniform steady state $H_s(x)$, we numerically calculate the dominant eigenvalues of the eigenproblem (28) with $\ell^* = 3.005$ and $0 \leq \beta \leq 0.002$. The eigenvalues are plotted in Fig. 14 (middle) showing that the three branches of eigenvalues come together at $\beta = 0$.

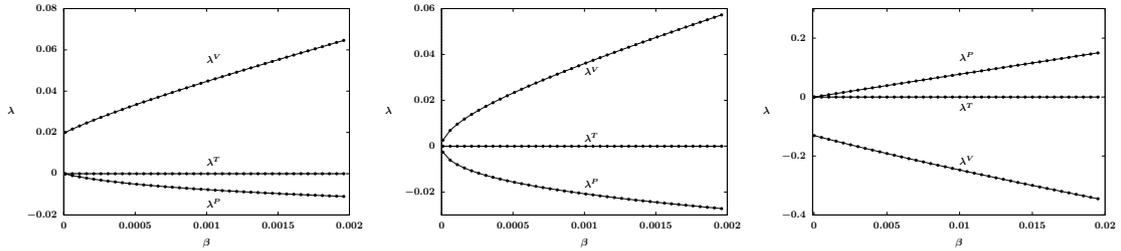

FIG. 14. Plots of the largest three eigenvalues of (28) over a range of $\beta$ with the period $\ell$ satisfying (left) $\ell = 2.9 < \ell^*$, (middle) $\ell = 3.005 = \ell^*$ and (right) $\ell = 3.2 > \ell^*$. The critical period is $\ell^* \approx 3.005$ for $P_0 = 0.5$.

For $\ell > \ell^*$, the dominant eigenvalues of the $\ell$-periodic steady state $H_s(x)$ are plotted against $\beta$ in Fig. 14 (right). It shows that in the limit $\beta \to 0$, the non-uniform steady state $H_s(x)$ is unstable to the $\Psi^P$ mode with $\lambda^P \sim \lambda_1^P \beta$ as $\beta \to 0$ where $\lambda_1^P > 0$. While for $\ell_s < \ell < \ell^*$, the plot in Fig. 14 (left) indicates that the unstable eigenvalue $\lambda^V$ takes the form $\lambda^V \sim \lambda_0^V + \lambda_1^V \beta$ as $\beta \to 0$ where $\lambda_0^V, \lambda_1^V > 0$.

For solutions with multiple droplets, additional unstable zero-mean coarsening modes that account for symmetry breaking and droplet merging coexist with the dominant modes described above [21]. We will consider the simplest case for steady states with two droplets where there is only one symmetry breaking mode $\Psi^B$, and one droplet merging mode $\Psi^M$.

## VI. NUMERICAL SIMULATIONS

To understand the influence of the stability of the non-uniform equilibria, we focus on the effects of the unstable eigenmodes of $H_s(x)$ based on PDE simulations of (1) starting from initial conditions of the form

$$h_0(x) = H_s(x) + \delta\Psi(x), \tag{44}$$

where $\Psi(x)$ is a normalized eigenmode associated with a positive eigenvalue $\lambda$ and $\|\Psi\|_2 = 1$, $\Psi(0) \geq 0$ and $|\delta| \ll 1$. Since it is assumed that the solution takes the form of $h(x,t) \sim H_s(x) + \delta\Psi(x)e^{\lambda t}$ when it is close to the steady state, the growth or decay of the perturbation in time can be quantified by

$$\|h(x,t) - H_s(x)\|_2 = \delta\|\Psi\|_2 e^{\lambda t} + O(\delta^2).$$

Unless explained otherwise, for all the simulations in this section, we keep $\beta = 0.001$ and $P_0 = 0.5$.

We begin by studying the dynamics from a single drop steady state $H_s(x)$. We use a numerical example to illustrate how the non-conservative eigenfunctions are tied to the evaporation dynamics. With $\ell = L = 3.0$ satisfying $\ell_s < \ell < \ell^*$, the eigenproblem for the non-uniform steady state $H_s$ has only one positive eigenvalue $\lambda^V = 1.44 \times 10^{-4}$. Starting from the initial data $h_0(x) = H_s - 0.01\Psi^V$ evaporation occurs and the solution eventually converges to the stable spatially-uniform steady state $h = H_m$ (see Fig. 15). Although the initial condition has average height $\langle h_0 \rangle > H_c$, suggestive of film-wise condensation based (13a), the strong influence of the spatial profile leads to drop-wise evaporation. In the early stage, Fig. 15 (left) shows that the spatial variations increase as the peak of the solution increases while minima decrease, and the growth rate of $\|h - H_s\|_2$ is consistent with the predicted growth rate given by the unstable eigenvalue $\lambda^V$ (see Fig. 16). In the later stage the convergence rate to the uniform steady state $h = H_m$ agrees with the analytical prediction from (17) with the wavenumber $k = 0$, $\lambda^m = -\beta\Pi'(H_m)/(H_m + K_0) = -0.28679$. If the simulation starts with $h_0(x) = H_s + 0.01\Psi^V$ instead, then condensation with decreasing spatial variations occurs eventually leading to flooding (23).

Now we study the more complicated dynamics that can occur with multiple droplets. For simplicity only two droplet steady states will be investigated. We consider the non-uniform steady state $H_{s,2}(x)$ with the period $\ell = 3.5 > \ell^*$ on $0 \leq x \leq 7$. By numerically solving the eigenproblem (28), we obtain three unstable eigenmodes (see Fig. 17) in addition to the translational zero eigenmode (not shown).

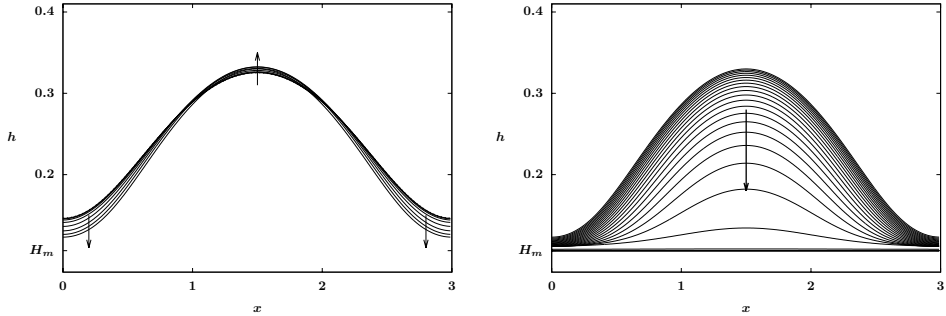

FIG. 15. (Left) Early stage evolution ($0 \leq t < 250$) and (right) the later stage ($t > 250$) of evaporation with initial condition close to non-uniform steady state solution $H_s$ for a PDE simulation starting from the initial data $h_0(x) = H_s - 0.0001\Psi^V$ with domain size $L = \ell = 3$.

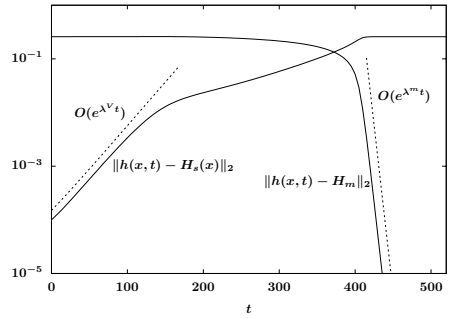

FIG. 16. The growth and decay rates of the $L^2$-norms between the PDE solution shown in Fig. 15 and the steady states $H_s(x)$ and $H_m$, showing the solution evolves from the $H_s(x)$ state towards $H_m$.

The eigenvalue $\lambda^B = 0.00736$ corresponds to the coarsening mode $\Psi^B$ that leads to symmetry breaking, $\lambda^M = 0.00611$ corresponds to the eigenmode $\Psi^M$ that yields droplet merging, and $\lambda^P = 0.00528$ is associated with the mode $\Psi^P$ which corresponds to evaporation or condensing effects. Among the three unstable eigenmodes, the only mode with non-zero mean is $\Psi^P$ that yields evaporative effects, and the other two eigenmodes govern the dynamics that conserve mass.

The set of PDE simulations in Fig. 18 demonstrate the effects of these eigenmodes as the initial conditions (44) with perturbation amplitude $\delta = 0.1$. With $\Psi(x) = \Psi^B$, initial perturbations lead to mass increase of the left droplet while the droplet on the right shrinks in time, breaking the symmetry of the original state (see Fig. 18 (a)). For $\Psi(x) = \Psi^M$ as the perturbation, the two droplets with the period $\ell = 3.5$ merge into one larger droplet with the period $\ell = 7$ (see Fig. 18 (b)). These two examples are similar to the coarsening dynamics observed in conservative thin film models [21]. In Fig. 18 (c) we show the results for the perturbation $\Psi = \Psi^P$ where initially spatial variations decrease. The solution eventually evolves into a spatially-uniform film which grows indefinitely in time. This flooding dynamics will also occur at later times for the simulations from Fig. 18 (a, b) with $\bar{h}(t) = O(t^{1/2})$ following (23), similar to Fig. 1.

The evolving average film thickness $\langle h \rangle$ of the three simulations are plotted in Fig. 18 (d) showing that among the three eigenmodes $\Psi^P$ yields the fastest condensation, while the dynamics driven by the two zero-mean modes are delayed by the early stage coarsening. Solutions from all these simulations are approaching flat films governed by large thickness limit (13a). Both Fig. 18 (a, b) proceed by coarsening to a growing single drop then leading to flooding. They differ in the position and formation time of that drop, but the flooding behavior is the same except for a time shift of $\Delta t = 217$ between the two.

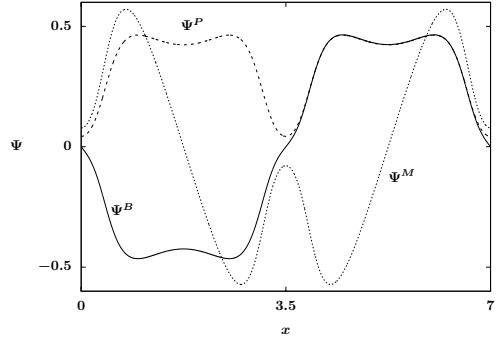

FIG. 17. Unstable eigenmodes of the non-uniform steady state $H_{s,2}(x)$ with two zero-mean modes $\Psi^B$, $\Psi^M$ and a non-zero-mean mode $\Psi^P$ where the domain size $L = 2\ell = 7$ and the other system parameters are identical to those in Fig. 15.

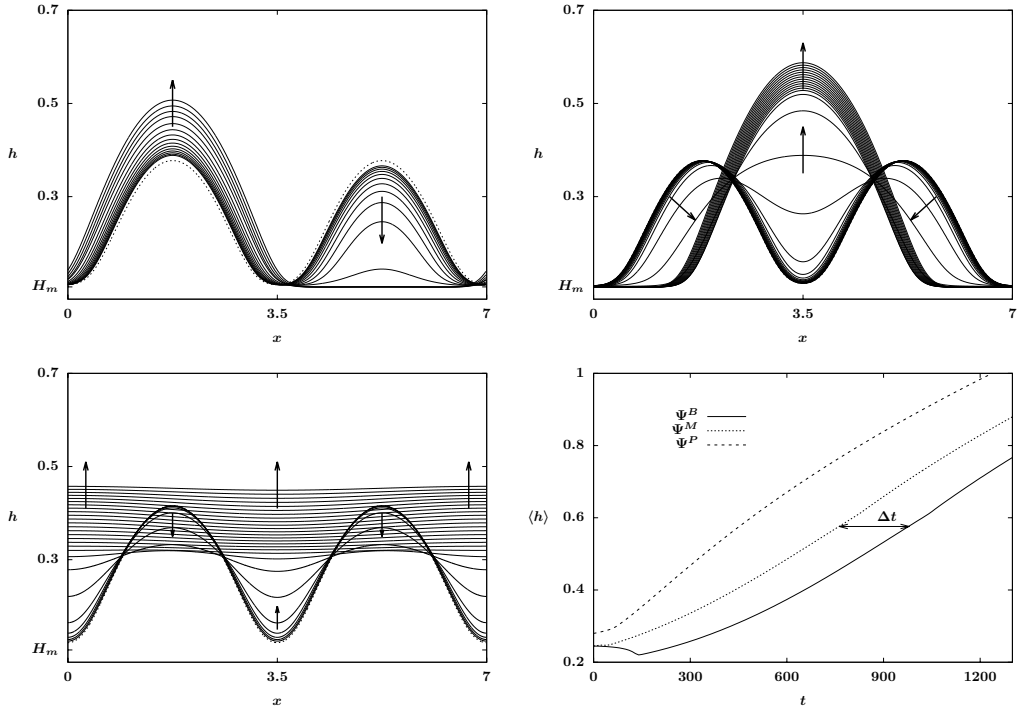

FIG. 18. PDE simulations of two droplets on $0 \leq x \leq 7$ starting from $h_0(x) = H_{s,2}(x) + 0.1\Psi$ showing (a) symmetry breaking with $\Psi = \Psi^B$, (b) droplet merging with $\Psi = \Psi^M$ and (c) flattening with $\Psi = \Psi^P$. The plot in (d) shows the corresponding changes of $\langle h \rangle$ in time, where the long-time behavior of the solutions perturbed by the $\Psi^B$ mode is identical to that of the $\Psi^M$ mode shifted by $\Delta t = 217$.

Fig. 19 gives a transition diagram that summarizes the relations between the instabilities and dynamics of the coexisting states studied above in a small domain for $\ell > \ell^*$. Droplet steady states $H_{s,1}$ and $H_{s,2}$ perturbed by the eigenmode $\Psi^P$ either converge to the constant equilibrium $H_m$ (see

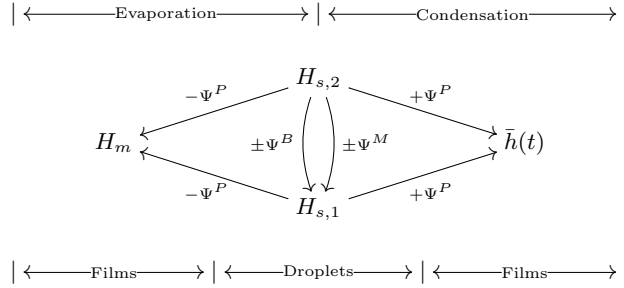

FIG. 19. A transition state diagram connecting linearized dynamics in the small domain case with $\ell < \ell^*$

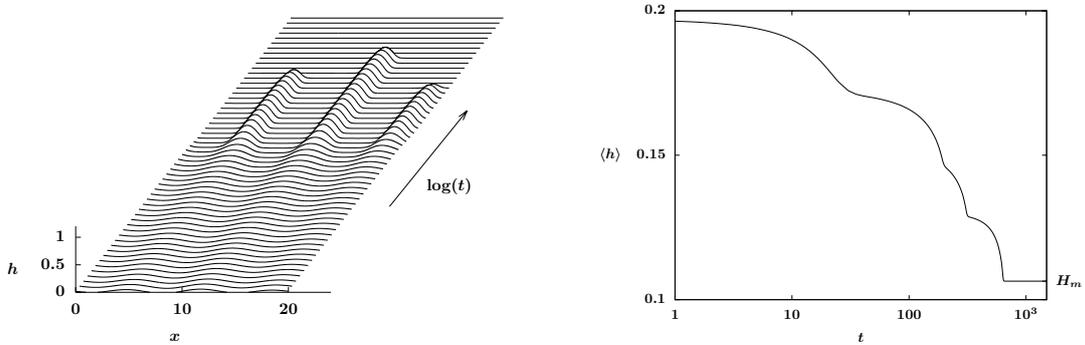

FIG. 20. (Left) PDE simulation starting from initial data $h_0(x) = 0.197 - 0.03\sin(6\pi x/L) + 0.01\sin(4\pi x/L)$ for the model (1) and (right) the corresponding monotonically decreasing average thickness in time.

Fig. 15) or exhibit film-wise condensation (see Fig. 18 (c)). The two coarsening modes $\Psi^B$ and $\Psi^M$ account for the transition from $H_{s,2}$ to $H_{s,1}$ up to a spatial translation (see Fig. 18 (a), b)). For droplets states with periods in the range of $\ell_s < \ell < \ell^*$, the stability of some of the eigenmodes can change, and the transition from droplets to flat films can be induced by the $\Psi^V$ mode. While film-wise dynamics with an increasing or decreasing total mass are relatively straightforward to understand (see section III), whether droplet states and coarsening dynamics will lead to evaporation or condensation is more difficult to identify.

With a larger domain size, more complicated dynamics are expected and need further investigation. With more droplet states involved, more unstable coarsening and flattening modes will appear in the corresponding transition diagram. For instance, it is shown in Fig. 20 that making a small perturbation (changing $\langle h_0 \rangle$) to the initial conditions used in (1) can lead to very different droplet dynamics than presented earlier in Fig. 1. Instead of drop-wise and film-wise condensation, the three isolated droplets all collapse for long times, indicating that the dynamics are sensitive to the initial conditions.

Motivated by the different dynamics shown in Fig. 1 and Fig. 20, we will now more systematically explore the dependence of evaporation versus condensation on initial conditions and the $\beta$ value. The plot in Fig. 21 shows simulation results starting from initial conditions

$$h_0(x) = \bar{h}_0 - 0.03\sin(6\pi x/L) + 0.01\sin(4\pi x/L)$$

over a range of values for $\bar{h}_0$ on the domain $0 \le x \le L$ with $L = 20$. We also use a range of values for $\beta$ in (1) to examine the impacts of weak versus strong evaporation relative to dewetting effects.

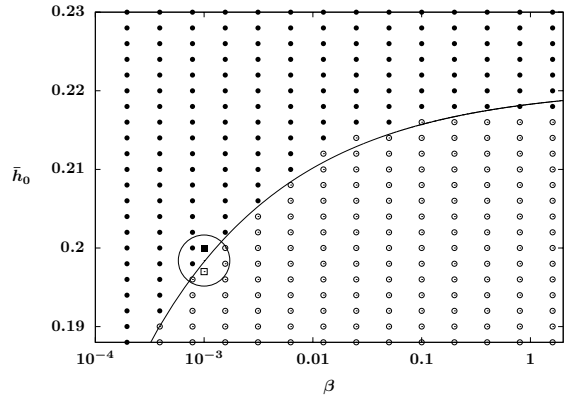

FIG. 21. A plot showing the influences of $\beta$ and the initial film height to evaporation (denoted by open circles) or condensation (denoted by closed circles) dynamics. The $(\beta, \bar{h})$ values used in Fig. 1 and Fig. 20 are labeled by closed and open squares, respectively.

For each simulation with a given $(\beta, \bar{h})$, whether long time evaporation or condensation is observed is labeled by open or closed circles, respectively in Fig. 21. For large $\bar{h}$ the dynamics rapidly converges to film-wise condensation, while sufficiently small $\bar{h}$ leads to evaporation. For a given $\beta$, there is an intermediate range of $\bar{h}$ at which we observe early stage coarsening dynamics followed by drop-wise evaporation or condensation dynamics, similar to the evolutions shown in Fig. 1 and Fig. 20. We use a simple curve to indicate the separate regions of observed evaporation and condensation dynamics. We suspect that the separation curve approaches $H_c$ as $\beta \to \infty$.

## VII. DISCUSSION

We have observed interesting transient pattern formation and interfacial instabilities of volatile thin fluid films from model (1) in the weak evaporation limit. In a critical range of pressures, we have shown that transitions between film-wise and drop-wise dynamics can yield various phase changes.

Unlike the conservative thin film equation (2) (equivalent to $\beta = 0$ in (1)) where a continuous family of non-uniform steady states with a constant pressure $p \equiv c$ exist in the system, the steady state solutions of (1) correspond to a single value of the pressure determined by the liquid-vapor balance at the imposed temperature. While models for coarsening dynamics in (2) have been constructed [22], because of the difference in the set of steady states there are still open questions on how to formulate similar reduced models for volatile films.

While the focus of this paper has been on weak evaporative effects, we are also interested in transient patterns under strong evaporative influences (with $\beta$ large). In particular, we would like to better understand the interaction between droplets coarsening and volatility, and identify the behavior of the evaporation/condensation dividing line in Fig. 21 in the large $\beta$ limit.

In this paper the dynamics of a two-dimensional evaporative/condensing thin film model have been demonstrated, and we expect some of the results to be naturally extended to the three-dimensional case. Furthermore, the current nonlinear analysis of equilibrium stability is limited to a relatively small system. For a larger scale domain more complicated and interesting pattern formation will arise and need further analysis.

We are also interested to further study the influences of other physical effects like vapor recoil, gravity and Marangoni effects to the pattern formation of volatile thin films [6, 7, 12, 37]. For instance, in [17] destabilizing gravitation and stabilizing surface tension are considered for a film under a cooled horizontal plate. In [37] Panzarella et al. incorporated buoyancy, capillary and evaporative effects to study a horizontal film boiling model.

**Appendix A: Positivity of solutions with $\beta \geq 0$**

In this section we show that for $\beta \geq 0$ the evaporative mass flux can not cause rupture singularities, and the solution stays positive.

The proof of this result depends on the properties of the disjoining pressure $\Pi(h)$ and the corresponding potential $U(h) = \int \Pi(h) \, dh$. From (7) we have $\lim_{h \to 0} U(h) = \infty$, and

$$\lim_{h \to \infty} U(h) = \begin{cases} -\infty & P_0 > 0 \\ 0 & P_0 = 0 \\ +\infty & P_0 < 0 \end{cases},$$

where the borderline case is consistent with the properties of potential function used in mass-conserving thin film equations in [11, 22]. The proof of this statement for $P_0 < 0$ is similar to the arguments from the paper [11]. For $P_0 > 0$, we use an alternative approach to show the regularity of the solution.

**Theorem 1.** *If $\beta \geq 0$ and initial data satisfies $h_0 > 0$, $h_0 \in H^1([0, L])$ and $\mathcal{E}[h_0] < \infty$, then a unique positive smooth solution to equation (1) exists for all $t > 0$.*

*Proof.* For $\beta \geq 0$ and $h > 0$, the functional (8) is monotonically dissipating. Using this fact, with the positive initial data, it suffices to derive a priori pointwise upper and lower bounds for the solution. Then the uniform parabolicity ensures that the solution is smooth and unique.

Since the functional is monotonically decreasing, at any time $T > 0$,

$$\frac{1}{2} \int_0^L \left| \frac{\partial h}{\partial x}(T) \right|^2 \, dx \leq \mathcal{E}[h_0] - \int_0^L U(h(T)) \, dx. \tag{A1}$$

For $P_0 \leq 0$, we directly observe that $-U(h)$ has an a priori upper bound independent of $h$, implying that $\int |\partial_x h(x, T)|^2 \, dx$ is bounded. For $P_0 > 0$, while $-U(h) \to \infty$ as $h \to \infty$, based on the form of $U(h)$ in (7) we notice that $-U(h)$ has an upper bound

$$-U(h) = \left( \frac{\epsilon^2}{2h^2} - \frac{\epsilon^3}{3h^3} \right) + P_0 h \leq \frac{1}{6} + P_0 h. \tag{A2}$$

Then we consider the estimate for $\int |\partial_x h(x, T)|^2 \, dx$. Using the Cauchy Schwarz inequality, one obtains that for any $0 \leq x, y \leq L$,

$$|h(y) - h(x)| = \left| \int_x^y \partial_s h(s) \, ds \right| \leq \int_0^L |\partial_s h(s)| \, ds \leq L^{\frac{1}{2}} \left( \int_0^L |\partial_s h(s)|^2 \, ds \right)^{\frac{1}{2}}. \tag{A3}$$

Suppose that at time $t = T$ the solution $h(x, T)$ attains its minimum $h_{\min}$ at $x = x_0$, then setting $y = x_0$ in (A3) yields

$$\int_0^L h(x, T) \, dx \leq \int_0^L |h_{\min} + (h(x, T) - h_{\min})| \, dx \leq |h_{\min}|L + L^{\frac{3}{2}} \left( \int_0^L |\partial_x h(x, t)|^2 \, dx \right)^{\frac{1}{2}}. \tag{A4}$$

Using (A2) and (A4) we get

$$- \int_0^L U(h(T)) \, dx \leq \frac{1}{6}L + P_0|h_{\min}|L + P_0 L^{\frac{3}{2}} \left( \int_0^L |\partial_x h(x, t)|^2 \, dx \right)^{\frac{1}{2}}. \tag{A5}$$

Combining (A5) and (A1) leads to

$$\frac{1}{2} \|\partial_x h(T)\|_{L^2}^2 - P_0 L^{\frac{3}{2}} \|\partial_x h(T)\|_{L^2} \leq \mathcal{E}[h_0] + \frac{1}{6}L + P_0|h_{\min}|L. \tag{A6}$$

Hence $h(x, T) \in H^1([0, L])$. Since the $H^1$-norm bounds the $L^\infty$-norm in one space dimension, $h(x, T)$ has an a priori pointwise upper bound. Then we derive an a priori pointwise lower bound for $h(x, T)$. First, from (A1) we note that

$$\int_0^L U(h(x, T)) \; dx < \mathcal{E}[h_0] \tag{A7}$$

By the Sobolev embedding theorem, for some constant $C_1$ we have $\|h(x, T)\|_{C^{0,1/2}} \le C_1 \|h(x, T)\|_{H^1}$. Again suppose that $h(x, T)$ attains its minimum $h_{\min}$ at $x = x_0$. By Hölder inequality, for some constant $C_h$ we obtain $h(x) \le h_{\min} + C_h |x - x_0|^{1/2}$. So for some constants $C_2(\epsilon) > 0$ and $C_3 > 0$, we have

$$\int_0^L U(h(x, T)) \; dx \ge \int_0^L \frac{C_2(\epsilon)}{h(x, T)^3} \; dx - C_3 \ge \int_0^L \frac{C_2(\epsilon)}{(h_{\min} + C_h |x - x_0|^{1/2})^3} \; dx - C_3, \tag{A8}$$

and

$$\int_0^L \frac{C_2(\epsilon)}{(h_{\min} + C_h |x - x_0|^{1/2})^3} \; dx \ge \int_{x_0 - (h_{\min}/C_h)^2}^{x_0 + (h_{\min}/C_h)^2} \frac{C_2(\epsilon)}{(2h_{\min})^3} \; dx = \frac{C_2(\epsilon)}{4 C_h^2 h_{\min}}. \tag{A9}$$

By (A7), (A8) and (A9), the solution cannot go below a positive threshold at any time $T > 0$. This completes our proof. $\qquad \square$

*Remarks:* When $\beta < 0$, the positivity of the solution is no longer guaranteed. Since the energy functional (8) is not Lyapunov any more, the energy argument presented in the proof does not apply to this case. In fact, finite-time rupture may occur and the PDE may break down past a critical time when specific values of system parameters are chosen; see [24].

---

[1] V. S. Ajaev. Evolution of dry patches in evaporating liquid films. *Physical Review E*, 72(3):031605, 2005.

[2] V. S. Ajaev. Spreading of thin volatile liquid droplets on uniformly heated surfaces. *Journal of Fluid Mechanics*, 528:279–296, 2005.

[3] V. S. Ajaev and G. M. Homsy. Steady vapor bubbles in rectangular microchannels. *Journal of colloid and interface science*, 240(1):259–271, 2001.

[4] V. S. Ajaev and O. A. Kabov. Heat and mass transfer near contact lines on heated surfaces. *International Journal of Heat and Mass Transfer*, 108:918–932, 2017.

[5] V. S. Ajaev, J. Klentzman, T. Gambaryan-Roisman, and P. Stephan. Fingering instability of partially wetting evaporating liquids. *J. Engrg. Math.*, 73:31–38, 2012.

[6] A. Amini and G. M. Homsy. Evaporation of liquid droplets on solid substrates. i. flat substrate with pinned or moving contact line. *Physical Review Fluids*, 2(4):043603, 2017.

[7] A. Amini and G. M. Homsy. Evaporation of liquid droplets on solid substrates. ii. periodic substrates with moving contact lines. *Physical Review Fluids*, 2(4):043604, 2017.

[8] D. M. Anderson and S. H. Davis. The spreading of volatile liquid droplets on heated surfaces. *Physics of Fluids*, 7(2):248–265, 1995.

[9] J. Becker, G. Grun, R. Seemann, H. Mantz, K. Jacobs, K. Mecke, and R. Blossey. Complex dewetting scenarios captured by thin-film models. *Nature Materials*, 2(1):59–63, 2003.

[10] A. J. Bernoff, A. L. Bertozzi, and T. P. Witelski. Axisymmetric surface diffusion: dynamics and stability of self-similar pinchoff. *Journal of Statistical Physics*, 93(3):725–776, 1998.

[11] A. L. Bertozzi, G. Grün, and T. P. Witelski. Dewetting films: bifurcations and concentrations. *Nonlinearity*, 14(6):1569, 2001.

[12] M. Bestehorn, A. Pototsky, and U. Thiele. 3D large scale Marangoni convection in liquid films. *The European Physical Journal B-Condensed Matter and Complex Systems*, 33(4):457–467, 2003.

[13] R. J. Braun. Dynamics of the tear film. *Annual Review of Fluid Mechanics*, 44:267–297, 2012.

[14] J. P. Burelbach, S. G. Bankoff, and S. H. Davis. Nonlinear stability of evaporating/condensing liquid films. *Journal of Fluid Mechanics*, 195:463–494, 1988.

[15] R. V. Craster and O. K. Matar. Dynamics and stability of thin liquid films. *Reviews of Modern Physics*, 81(3):1131, 2009.

[16] M. Cross and H. Greenside. *Pattern formation and dynamics in nonequilibrium systems*. Cambridge University Press, 2009.

[17] R. J. Deissler and A. Oron. Stable localized patterns in thin liquid films. *Physical review letters*, 68(19):2948, 1992.

[18] R. Enright, N. Miljkovic, J. L. Alvarado, K. Kim, and J. W. Rose. Dropwise condensation on micro-and nanostructured surfaces. *Nanoscale and Microscale Thermophysical Engineering*, 18(3):223–250, 2014.

[19] P. L. Evans, L. W. Schwartz, and R. V. Roy. A mathematical model for crater defect formation in a drying paint layer. *Journal of colloid and interface science*, 227(1):191–205, 2000.

[20] A. Ghatak, R. Khanna, and A. Sharma. Dynamics and morphology of holes in dewetting of thin films. *Journal of colloid and interface science*, 212(2):483–494, 1999.

[21] K. Glasner and T. Witelski. Collision versus collapse of droplets in coarsening of dewetting thin films. *Physica D: Nonlinear Phenomena*, 209(1):80–104, 2005.

[22] K. B. Glasner and T. P. Witelski. Coarsening dynamics of dewetting films. *Physical review E*, 67(1):016302, 2003.

[23] L. M. Hocking. The influence of intermolecular forces on thin fluid layers. *Physics of Fluids A*, 5(4):793–799, 1993.

[24] H. Ji and T. P. Witelski. Finite-time thin film rupture driven by modified evaporative loss. *Physica D*, 342:1–15, 2017.

[25] J. R. King, A. Muench, and B. A. Wagner. Linear stability analysis of a sharp-interface model for dewetting thin films. *Journal of Engineering Mathematics*, 63(2-4):177–195, 2009.

[26] R. S. Laugesen and M. C. Pugh. Linear stability of steady states for thin film and Cahn-Hilliard type equations. *Archive for rational mechanics and analysis*, 154(1):3–51, 2000.

[27] R. S. Laugesen and M. C. Pugh. Properties of steady states for thin film equations. *European Journal of Applied Mathematics*, 11(03):293–351, 2000.

[28] R. S. Laugesen and M. C. Pugh. Energy levels of steady states for thin-film-type equations. *Journal of Differential Equations*, 182(2):377–415, 2002.

[29] N. Miljkovic and E. N. Wang. Condensation heat transfer on superhydrophobic surfaces. *MRS Bulletin*, 38(05):397–406, 2013.

[30] A. Muench and B. Wagner. Impact of slippage on the morphology and stability of a dewetting rim. *Journal of Phyiscs-Condensed Matter*, 23(18, SI), 2011.

[31] T. G. Myers. Thin films with high surface tension. *SIAM Review*, 40(3):441–462, 1998.

[32] J. Ockendon. *Viscous flow*, volume 13. Cambridge University Press, 1995.

[33] A. Oron and S. G. Bankoff. Dewetting of a heated surface by an evaporating liquid film under conjoining/disjoining pressures. *Journal of colloid and interface science*, 218(1):152–166, 1999.

[34] A. Oron and S. G. Bankoff. Dynamics of a condensing liquid film under conjoining/disjoining pressures. *Physics of Fluids*, 13(5):1107–1117, 2001.

[35] A. Oron, S. H. Davis, and S. G. Bankoff. Long-scale evolution of thin liquid films. *Reviews of Modern Physics*, 69(3):931, 1997.

[36] A. Padmakar, K. Kargupta, and A. Sharma. Instability and dewetting of evaporating thin water films on partially and completely wettable substrates. *The Journal of chemical physics*, 110(3):1735–1744, 1999.

[37] C. H. Panzarella, S. H. Davis, and S. G. Bankoff. Nonlinear dynamics in horizontal film boiling. *Journal of Fluid Mechanics*, 402:163–194, 2000.

[38] G. Reiter. Unstable thin polymer-films - rupture and dewetting processes. *Langmuir*, 9(5):1344–1351, 1993.

[39] J. Rose. Proc. inst. mech. eng. part a-j. *Power Energy*, 216:115, 2002.

[40] K. Rykaczewski, A. T. Paxson, M. Staymates, M. L. Walker, X. Sun, S. Anand, S. Srinivasan, G. H. McKinley, J. Chinn, J. H. J. Scott, et al. Dropwise condensation of low surface tension fluids on omniphobic surfaces. *Scientific reports*, 4, 2014.

[41] R. Schaaf. Global behaviour of solution branches for some neumann problems depending on one or several parameters. *Journal für die reine und angewandte Mathematik*, 346:1–31, 1984.

[42] L. W. Schwartz, R. V. Roy, R. R. Eley, and S. Petrash. Dewetting patterns in a drying liquid film. *Journal of colloid and interface science*, 234(2):363–374, 2001.

[43] A. Sharma. Equilibrium and dynamics of evaporating or condensing thin fluid domains: thin film stability and heterogeneous nucleation. *Langmuir*, 14(17):4915–4928, 1998.

[44] A. Sharma and R. Khanna. Pattern formation in unstable thin liquid films. *Physical Review Letters*, 81(16):3463, 1998.

[45] A. Sharma and G. Reiter. Instability of thin polymer films on coated substrates: Rupture, dewetting,

and drop formation. *Journal of Colloid and Interface Science*, 178(2):383–399, 1996.

[46] O. E. Shklyaev and E. Fried. Stability of an evaporating thin liquid film. *Journal of Fluid Mechanics*, 584:157–183, 2007.

[47] E. Sultan, A. Boudaoud, and M. Ben Amar. Evaporation of a thin film: diffusion of the vapour and Marangoni instabilities. *Journal of Fluid Mechanics*, 543:183–202, 2005.

[48] U. Thiele. Thin film evolution equations from (evaporating) dewetting liquid layers to epitaxial growth. *Journal of Physics: Condensed Matter*, 22(8):084019, 2010.

[49] E. M. Tian and D. J. Wollkind. A nonlinear stability analysis of pattern formation in thin liquid films. *Interfaces and Free Boundaries*, 5(1):1–25, 2003.

[50] D. Todorova, U. Thiele, and L. M. Pismen. The relation of steady evaporating drops fed by an influx and freely evaporating drops. *Journal of Engineering Mathematics*, 73(1):17–30, 2012.

[51] P. C. Wayner. Intermolecular forces in phase-change heat transfer: 1998 Kern award review. *AIChE Journal*, 45(10):2055–2066, 1999.

[52] R. Wen, Z. Lan, B. Peng, W. Xu, R. Yang, and X. Ma. Wetting transition of condensed droplets on nanostructured superhydrophobic surfaces: Coordination of surface properties and condensing conditions. *ACS Applied Materials & Interfaces*, 9(15):13770–13777, 2017.

[53] R. Wen, Q. Li, J. Wu, G. Wu, W. Wang, Y. Chen, X. Ma, D. Zhao, and R. Yang. Hydrophobic copper nanowires for enhancing condensation heat transfer. *Nano Energy*, 33:177–183, 2017.

[54] M. B. Williams and S. H. Davis. Nonlinear theory of film rupture. *Journal of colloid and interface science*, 90(1):220–228, 1982.

[55] T. P. Witelski and A. J. Bernoff. Dynamics of three-dimensional thin film rupture. *Physica D*, 147(1):155–176, 2000.

[56] W. W. Zhang and J. R. Lister. Similarity solutions for van der Waals rupture of a thin film on a solid substrate. *Phys. Fluids*, 11(9):2454–2462, 1999.